# From Research Gaps to Theoretical Opportunities: Theory-Oriented GenAI for Research Opportunity Evaluation

**Jujun Huang**
School of Management, Binghamton University, State University of New York
Binghamton, New York 13902
jhuang83@binghamton.edu

**Shun Cao**
School of Management, Binghamton University, State University of New York
Binghamton, New York 13902
scao12@binghamton.edu

Corresponding author: Shun Cao, scao12@binghamton.edu

**Abstract:** Generative AI (GenAI) can explore large bodies of literature and generate plausible research ideas, but identifying what is missing, understudied, contradictory, or potentially connected does not by itself reveal where theory should advance. We develop a theory-oriented agentic AI system that helps researchers identify potential theorizing opportunities by incorporating established theorizing approaches into literature exploration and evaluation. The system operates through three stages. Stage 1 expands the theoretical search space and constructs a provisional Candidate Knowledge Graph. Stage 2 independently reconstructs what the literature supports through source-grounded evidence extraction and theory-state reconstruction. Stage 3 evaluates the reconstructed knowledge state to determine whether an unresolved configuration warrants theory development or another research action and, when theory development is warranted, which theorizing approach is appropriate. We demonstrate the system through an end-to-end analysis of human oversight of agentic AI systems in organizations. The analysis shows that literature gaps alone are insufficient for identifying theoretical opportunities. For example, transparency → trust is routed to mechanism-based theorizing because the relationship is repeatedly documented in prior studies, while the generative mechanism explaining how transparency shapes trust remains insufficiently specified. By combining large-scale literature processing, structured knowledge representation, and theorizing-guided diagnosis, the system serves as a theory-oriented research assistant that supports researchers in identifying theoretically meaningful directions for subsequent research.

## 1. Introduction

Generative AI (GenAI) is increasingly used to support scholarly ideation and knowledge exploration. Existing systems can scan large literatures, generate research directions and hypotheses, and identify apparent gaps or potentially novel connections (Wang et al. 2024; Yang et al. 2024; Baek et al. 2025; Gopal et al. 2025). These capabilities expand the research space scholars can explore. Yet in theory-intensive fields such as information systems and management, identifying what is missing, understudied, contradictory, or potentially connected does not establish a theoretically meaningful research opportunity. The more consequential question is whether an unresolved pattern constitutes a problem for existing theoretical understanding and, if so, whether and how theory should be developed. Expanding the searchable knowledge space therefore does not necessarily improve the ability to determine where theory should move next.

This limitation becomes more important as GenAI assumes a more active role in knowledge production. Gopal et al. (2025) describes the automation of invention as a shift in which AI increasingly participates in loops connecting idea generation, research design, execution, interpretation, and evaluation. For theory-oriented research, the harder problem is therefore no longer generating plausible possibilities, but deciding what scholarly action the accumulated evidence justifies. A gap may call for more evidence, replication, measurement work, conceptual clarification, or theory development. We address this problem by incorporating established theorizing approaches into a system that separates possibility, evidence, and justified research action.

Existing AI-supported research systems offer important capabilities for gap identification, scientific ideation, and literature-grounded knowledge development (Schryen et al. 2025; Wang et al. 2024; Yang et al. 2024; Jansen et al. 2026), but provide limited guidance for determining whether a candidate opportunity is theoretically consequential. An unexplored construct combination, understudied context, contradiction, or missing connection may warrant investigation, but it may also reflect problems of measurement, evidence, research design, retrieval coverage, or conceptual clarity rather than a need for theory development. We therefore treat opportunity identification as a theory-grounded research-action problem and define the theoretical opportunity space as unresolved configurations for which theory development is the justified research response. A theoretical opportunity exists when theory development is warranted, at least one theorizing approach is appropriate for the diagnosed theoretical problem, and prior-art assessment indicates that the problem remains unresolved. This leads to our central research question: **How can a GenAI-supported scholarly system determine when theory development is warranted and identify an appropriate theorizing route?**

We address this question by treating theoretical opportunity not as a property of a research gap, but as an evidence-grounded research-action judgment that emerges only after the relevant knowledge state has been reconstructed and diagnosed. To operationalize this perspective, we develop a theory-oriented agentic research system built around deductive theory elaboration (Fisher and Aguinis 2017; Paré et al. 2026), mechanism-based theorizing (Hedström and Ylikoski 2010; Siponen et al. 2025), boundary-condition theorizing (Whetten 1989; Sun and Rivard 2026), and context-based theorizing (Johns 2006; Vaast 2026). These approaches provide distinct criteria for extending, explaining, bounding, or developing theory from context. The system does not assume that every unresolved pattern or research gap requires theory development. It reconstructs the accumulated theoretical and evidentiary state and assesses whether theory development is warranted; only configurations that pass this judgment are evaluated against the eligibility conditions of the theorizing approaches. The architecture implements this logic through minimum necessary agency (Pacella et al. 2026), epistemic separation (Nguyen 2024), specialized reasoning with controlled integration (Li et al. 2023; Chen et al. 2024; Prasad et al. 2024), and verification before epistemic escalation with bounded autonomy and human governance (Zhu et al. 2026; Nguyen et al. 2026).

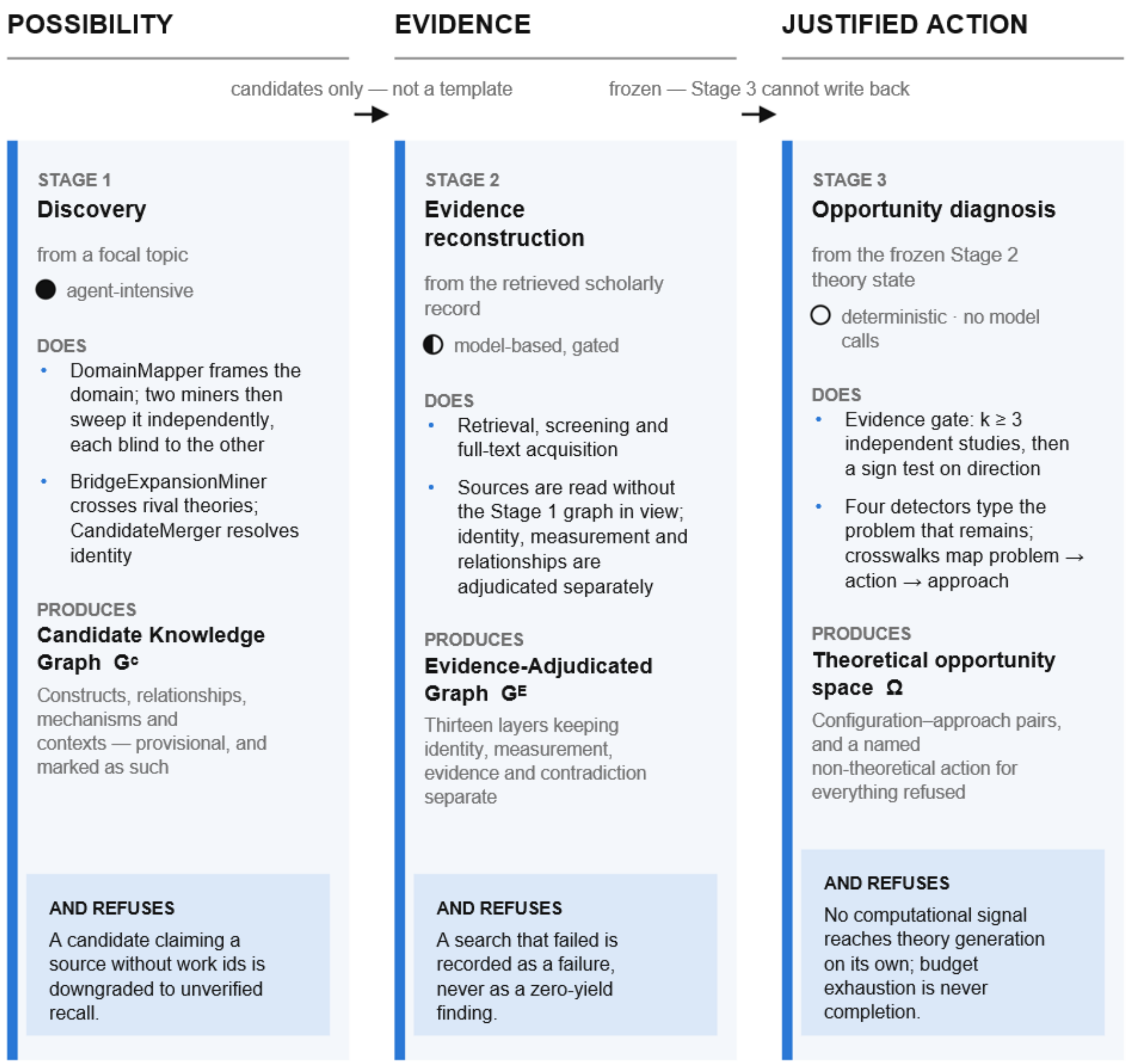

**Figure 1.** The Three-Stage Platform and Its Epistemic States, and What Crosses Between Them. Notes. The state above each column is the epistemic status of that stage's output; the label over each gap is the only condition under which material may cross it, and the struck return arrow marks a path the architecture does not provide. The disc beside each stage name is model discretion: filled for agent-intensive, half-filled for model-based but gated, open for deterministic with no model calls. The figure carries no run-specific counts; Section 4 reports the run.

The platform consists of three epistemically distinct stages summarized in Figure 1. Stage 1, the Discovery Stage, expands the theoretical search space and constructs a provisional Candidate Knowledge Graph of potentially relevant constructs, relationships, mechanisms, contexts, and other theoretical elements. Stage 2, the Evidence Reconstruction Stage, independently reads and adjudicates retrieved sources, resolves construct and theory identities, and reconstructs a multilayer Evidence-Adjudicated Knowledge Graph without using the Stage 1 graph as an extraction template. Stage 3, the Opportunity Diagnosis Stage, evaluates the reconstructed evidence state, determines whether an unresolved configuration warrants theory development or a non-theoretical research action, and, when appropriate, routes it to a theorizing approach. The stages thus separate possibility, evidence, and justified action rather than collapsing them into a single generative judgment. Figure 1 also records what each stage refuses. The separation between the three states is enforced by those refusals rather than asserted by the stage boundaries alone.

We demonstrate the three-stage process through an empirical run on the topic of "*human oversight of agentic AI systems in organizations*." Stage 1 generated a provisional Candidate Knowledge Graph containing 6,585 proposed relationships, treating these relationships as possibilities rather than established knowledge. Stage 2 independently reconstructed the scholarly record, yielding 14,214 literature-grounded relationships supported by source-grounded claims. Stage 3 then sharply narrowed this reconstructed knowledge state: only 28 relationships were supported by at least three independent studies, and only two—transparency → trust and explainability → trust—satisfied the directional-evidence criterion required for theoretical diagnosis. Across these two configurations, the system identified five typed theoretical problems: both exhibited a missing mechanism and a boundary failure, while transparency → trust additionally exhibited a misspecified relationship. Applying the fixed diagnosis–action–approach crosswalk yielded seven distinct configuration–approach pairs spanning all four focal theorizing approaches. These seven pairs therefore constitute the theory-development candidate set rather than confirmed theoretical opportunities because prior-art unresolvedness was not assessed in the current run. They are passed forward for human review and scholarly judgment before any theoretical opportunity or resulting contribution is ultimately accepted. Taken together, the run illustrates a theory-guided and governed agentic design in which AI agents help researchers identify theory-development candidates from large bodies of literature while leaving final theoretical judgment to human researchers.

This study makes three contributions. First, we reconceptualize AI-supported research opportunity identification by distinguishing a research gap from a theoretical opportunity. A theoretical opportunity is not established by novelty, absence, contradiction, or weak connection alone; it is an evidence-grounded research-action judgment that requires sufficient reason to treat theory development, rather than additional evidence, replication, measurement work, or conceptual clarification, as the appropriate response. Second, we show how established theorizing approaches can function as diagnostic logics for scholarly opportunity evaluation. Rather than using theorizing approaches merely as prompts for generating explanations after a gap has been identified, the system uses them to determine what form of theoretical development is appropriate for a diagnosed theoretical problem. Third, we develop a hybrid agentic–deterministic architecture that enforces epistemic separation between generated possibility, reconstructed evidence, and justified research action. The architecture allocates model discretion where open-ended interpretation is needed while using deterministic gates, refusal conditions, and human authority to govern epistemic escalation. Together, these contributions shift the objective of AI-supported scholarly discovery from generating more plausible research ideas toward making more disciplined judgments about which ideas warrant theoretical development.

## 2. Theoretical Background and Related Works

Our argument connects two bodies of research that have developed largely in parallel. Research on theorizing provides established logics for determining how theoretical explanations can be extended, refined, bounded, or developed. Research on GenAI-supported scholarship provides computational capabilities for searching large literatures, extracting structured knowledge, generating candidate connections, and coordinating complex analytical workflows. The first literature provides criteria for judging when and how theory should advance; the second provides tools for exploring and reconstructing the knowledge on which such judgments depend. We develop the theoretical basis for connecting these two streams below.

### 2.1. Theorizing Approaches for Theoretical Advancement

Our focus is not research opportunity identification in general, but opportunities for theoretical advancement. Theory explains phenomena through constructs, relationships, mechanisms, assumptions, and conditions that specify how and why outcomes occur. Advancing theory therefore requires more than adding an empirical association, studying an unexplored setting, or identifying a previously unexamined combination of constructs. It requires showing how an existing explanation should be refined, extended, qualified, or replaced.

Literature offers several established approaches for making such advances. We focus on four established approaches that provide distinct logics for how theoretical understanding can be advanced once the relevant theoretical problem has been identified: deductive theory elaboration, mechanism-based theorizing, boundary-condition theorizing, and context-based theorizing (Bacharach 1989; Whetten 1989; Bamberger 2008; Hedstrom and Swedberg 1998; Hedstrom and Ylikoski 2010; Fisher and Aguinis 2017; Pare et al. 2026). The four approaches are not intended to exhaust the IS theorizing repertoire. We select them because they represent distinct and sufficiently explicit logics for refining theoretical architecture, explaining mechanisms, specifying boundaries, and developing theory from context. They are not necessarily mapped one-to-one to theoretical problems: a configuration may present multiple problems, and a diagnosed problem may make more than one theorizing approach appropriate.

Deductive theory elaboration develops an existing theoretical explanation through conceptually justified changes to its internal architecture (Fisher and Aguinis 2017; Pare et al. 2026). The starting point is an identifiable theory or conceptual model that remains useful but contains constructs, relationships, assumptions, or boundaries that are incomplete or insufficiently specified. Elaboration may therefore involve refining a construct, decomposing an overly broad concept, adding a theoretically necessary relationship, or revising assumptions about when an explanation applies. Burton-Jones and Straub's (2006) reconceptualization of system usage illustrates this logic in IS. Rather than abandoning usage as a construct, they clarified its theoretical structure by distinguishing the user, system, and task elements relevant to a particular nomological network. Deductive elaboration is therefore appropriate when the central problem lies within the architecture of an otherwise viable theoretical explanation.

Mechanism-based theorizing addresses a different deficiency: an empirical phenomenon or relationship may be sufficiently established while the process that generates it remains underspecified. Mechanism-based explanation seeks to identify the relevant entities, activities, and interactions through which an outcome is produced rather than treating the association between an antecedent and an outcome as an explanation in itself (Machamer et al. 2000; Hedstrom and Ylikoski 2010; Siponen et al. 2025). Vayrynen et al. (2025), for example, explain industry-level digital transformation through a regulatory ambiguity mechanism connecting regulatory provisions, stakeholder interpretations, and technology-adoption decisions. The theoretical advance comes from opening the explanatory "black box." When theory development is warranted, mechanism-based theorizing is appropriate when the focal regularity is credible but existing theory does not adequately explain how or why it occurs.

Boundary-condition theorizing asks where an existing theoretical explanation does and does not apply. Theoretical claims necessarily depend on assumptions about actors, settings, technologies, time periods, and levels of analysis (Bacharach 1989; Whetten 1989; Busse et al. 2017). Evidence that a relationship changes systematically across such conditions may therefore reveal a limitation in explanatory scope rather than random heterogeneity. Boundary theorizing makes these conditions explicit and examines whether they change the strength, form, or even direction of a theoretical relationship. Recent IS scholarship similarly emphasizes that revising the boundary of a theory can itself constitute theoretical advancement when it changes the assumptions that delimit the theory's explanatory domain (Sun and Rivard 2026). This approach is appropriate when an explanation is otherwise viable but its reach is conditional rather than general.

Context-based theorizing gives context a still different role. Here, context is not merely a setting in which an established theory is retested; characteristics of the setting become part of the explanation itself (Johns 2006; Bamberger 2008; Whetten 2009; Hong et al. 2014; Vaast 2026). Close engagement with actors, practices, technologies, structures, or processes in a particular setting may reveal theoretical relationships or mechanisms that existing theories do not make visible. Lindberg et al. (2024), for example, use the practices of open-source software communities to develop the concept of discursive modulation and explain how discourse becomes embedded in software novelty and complexity. Context-based theorizing is therefore appropriate when features of the setting reveal explanatory structures that cannot be understood by simply transporting an existing theory into a new empirical domain. Table 1 summarizes the four approaches, the theoretical problem each addresses, and how each advances theory.

**Table 1.** Theorizing Approaches for Theoretical Advancement

| Theorizing approach | Core theoretical problem | How theory is advanced |
|---|---|---|
| Deductive theory elaboration | Existing theoretical architecture is useful but incomplete or imprecise | Refines constructs, relationships, assumptions, or boundaries within an existing explanation (Fisher and Aguinis 2017; Paré et al. 2026) |
| Mechanism-based theorizing | A phenomenon or relationship is established but its generative process is underspecified | Explains how and why the phenomenon is produced by specifying relevant entities, activities, and interactions (Hedstrom and Ylikoski 2010; Siponen et al. 2025) |
| Boundary-condition theorizing | An explanation appears to hold only under particular theoretically meaningful conditions | Clarifies where an explanation holds, changes, weakens, or fails (Whetten 1989; Sun and Rivard 2026) |
| Context-based theorizing | Features of a context reveal explanatory structures not adequately captured by existing theory | Develops theoretical insight from context rather than merely applying theory in another setting (Johns 2006; Hong et al. 2014; Vaast 2026) |

### 2.2. From AI-Supported Knowledge Exploration to Theoretical Opportunity Spaces

The GenAI has expanded the range of scholarly activities that can be computationally supported. Beyond drafting and summarization, recent systems participate in literature assessment, scientific ideation, hypothesis generation, and other upstream research activities. Gopal et al. (2025) characterize this broader development as the "automation of invention," in which AI participates in increasingly connected cycles of idea generation, analysis, interpretation, and evaluation. For theory-oriented research, the relevant implication is not simply that AI can generate more ideas. It is that AI can help scholars search and interpret bodies of knowledge that are difficult for an individual researcher to examine comprehensively. Several streams of work illustrate this potential. Research on GenAI-supported literature reviews considers activities ranging from synthesis and evidence aggregation to critique, theory building, gap identification, and research-agenda development (Schryen et al. 2025). Scientific ideation systems move further toward prospective knowledge development. SciMON searches prior work and generates research directions evaluated for novelty (Wang et al. 2024), while MOOSE combines scholarly and external knowledge to generate scientific hypotheses (Yang et al. 2024). Agentic research systems coordinate multiple research activities; ResearchAgent, for example, combines scholarly retrieval, knowledge structures, idea generation, and reviewing agents (Baek et al. 2025). Jansen et al. (2026) move closer to literature-grounded theory construction by synthesizing higher-level scientific regularities from collections of published studies.

These systems address important parts of scholarly discovery, but identifying a theoretical opportunity requires an additional judgment. A plausible hypothesis, an unexplored relationship, or a weakly connected research stream does not by itself show that theory requires development. The same apparent gap may arise because the evidence is sparse, constructs are defined differently across studies, measures are not comparable, studies address different levels or contexts, or the relevant literature has not been adequately retrieved. Theory-oriented opportunity identification therefore requires reconstruction of the accumulated knowledge state before deciding what kind of research action that state warrants. This distinction is central to the architecture developed in Section 3.

Recent advances in scientific information extraction, language-model reasoning, agentic workflows, and knowledge graphs make it increasingly feasible to reconstruct and explore large scholarly knowledge spaces (Luan et al. 2018; Ji et al. 2022; Yao et al. 2023a, 2023b; Dagdelen et al. 2024; Prasad et al. 2024). These capabilities can structure distributed claims, preserve intermediate analytical states, and coordinate distinct retrieval, interpretation, and validation tasks. What remains less developed is the subsequent theory-oriented judgment: whether an unresolved configuration represents a deficiency in theoretical understanding, whether theory development is the appropriate research action, and, if so, which theorizing logic fits the diagnosed problem. Table 2 positions our system around this distinction.

Table 2 highlights the analytical function that remains underdeveloped in existing AI-supported scholarly systems. Scientific ideation systems expand plausible research directions, agentic workflows coordinate research activities, and literature-grounded systems reconstruct or synthesize prior knowledge. Our focus is the subsequent judgment: whether an unresolved configuration constitutes a problem for theoretical understanding and, if so, what research action is justified. This judgment requires separating candidate possibilities from literature-supported knowledge and distinguishing theoretical deficiencies from problems that call for additional evidence, replication, measurement, conceptual clarification, or other non-theoretical actions. Section 3 operationalizes this distinction through the separation of discovery, evidence reconstruction, and opportunity diagnosis.

**Table 2.** AI-Supported Scholarly Research and Theory-Oriented Opportunity Identification

| Existing approach | Literature-grounded | Multi-stage reasoning | Theory-state reconstruction | Theory-development diagnosis | Theorizing guidance |
|---|---|---|---|---|---|
| Scientific ideation systems (SciMON; MOOSE) | Partial | Partial | No | No | Not explicit |
| Agentic research workflows (ResearchAgent) | Partial | Yes | No | No | Not explicit |
| Literature-grounded theory construction (Jansen et al. 2026) | Yes | Yes | Partial | No | Not explicit |
| Proposed theory-oriented agentic exploration | Yes | Yes | Yes | Yes | Yes |

*Note.* Entries describe designed capability, not comparative performance; no system in this table was benchmarked against another.

## 3. Design of the Agentic Research System

### 3.1. Conceptualizing the Theoretical Opportunity Space

We conceptualize the theoretical opportunity space as theoretically consequential unresolved configurations for which theory development is an appropriate research response. A missing relationship, understudied context, inconsistent finding, or potentially novel connection may warrant investigation, but does not by itself establish a need for theory development. Identifying a theoretical opportunity therefore requires an epistemic escalation from what may be relevant, to what the scholarly record supports, to what research action that evidence justifies. The three stages preserve these states separately so that generated possibility is not treated as evidence and evidentiary support is not automatically treated as a warrant for theory development.

Let $\mathcal{I}$ denote the initial information available for the focal research topic. Stage 1 uses $\mathcal{I}$ to construct a Candidate Knowledge Graph, $G^{\mathrm{c}} = F_1(\mathcal{I}) = (V^{\mathrm{c}}, E^{\mathrm{c}})$, where $V^{\mathrm{c}}$ consists of theoretically relevant entities, including constructs, theories, mechanisms, processes, contexts, boundary conditions, and measures, and $E^{\mathrm{c}}$ captures candidate relationships among these entities. Importantly, $G^{\mathrm{c}}$ is a provisional candidate representation rather than an evidentiary claim, which means that Stage 1 identifies what may warrant investigation, not what the literature establishes. Stage 2 independently reconstructs the scholarly record from the retrieved literature $\mathcal{L}$ as an Evidence-Adjudicated Knowledge Graph, $G^{\mathrm{E}} = F_2(\mathcal{L}) = \{G_l^E : \ell \in \Lambda\}$. Here, $\Lambda$ indexes the interconnected layers $G_l^E$ used to organize literature-derived evidence, claims, constructs and measures, relationships, and theory structures. $G^{\mathrm{E}}$ represents what the retrieved scholarly record supports under the review protocol, in contrast to the provisional candidate representation generated in Stage 1. The procedures used to construct and adjudicate these layers are described in Section 3.2.2.

After Stage 2 has reconstructed and internally resolved $G^E$, the platform reconciles it with $G^C$. Reconciliation distinguishes Stage 1 candidates for which literature support is located, candidates for which support is not located under the current review protocol, and literature-supported elements that Stage 1 did not anticipate. Node correspondence is based on substantive meaning, referent, and level of analysis rather than label similarity alone, and relationship correspondence is evaluated only after construct identities and relationship types have been resolved. Detailed reconciliation functions are reported in the Online Appendix.

The reconstructed evidence state and its reconciliation with Stage 1 identify signals for further evaluation in Stage 3. Computational patterns such as missing links, structural holes, or disconnected graph regions may trigger examination, but they cannot by themselves establish a theoretical opportunity. Signals that survive verification become focal configurations $z$. For each configuration, Stage 3 assembles the relevant evidence, theory versions, mechanisms, assumptions, boundaries, contexts, study designs, and quality information into a structured theory state $x(z)$. The full representation is reported in the Online Appendix.

Stage 3 first evaluates whether an identifiable theory adequately explains the focal configuration, considering its mechanisms, assumptions, boundaries, prior extensions, and relevant non-theoretical alternatives. It then diagnoses the source of any inadequacy, including whether another evidentiary or conceptual problem must be resolved before theory development can be justified. TheoryActionGate converts this diagnosis and the reconstructed knowledge state into a current theory-development warrant, $W(z) \in \{0, 1, \perp\}$. $W(z) = 1$ indicates that theory development is currently warranted, $W(z) = 0$ indicates that a non-theoretical research action is warranted, and $W(z) = \perp$ indicates that the available evidence is insufficient to determine the appropriate action. The complete diagnosis representation is reported in the Online Appendix.

| CONDITION | WHAT IT ASKS | IF IT REFUSES, THE ACTION IS |
|---|---|---|
| **k** evidence adequacy | Do at least k independent studies support the relationship? (k = 3 under the default profile) | gather more evidence first |
| **α** direction agreement | Do those studies agree on direction, by an exact sign test at level α? | replicate first, where they conflict |
| **Γ(z)** problem diagnosis | Which typed theoretical problem does the relationship present? Four detectors ask independently, and one configuration may present several. | extend scope, or no development justified |
| **W(z)** action warrant | Is theory development the justified action, or is one of the nine non-theoretical actions? | the named non-theoretical action |
| **$I_a(z)$** approach eligibility | Which theorizing approaches does that diagnosis license (Figure 7a)? Every eligible approach is retained. | — no configuration is refused here |
| **$U_a(z)$** prior-art unresolvedness | Has prior theory already addressed this problem for this relationship? | cite the existing account; no new theory is warranted |

↓

**Ω the theoretical opportunity space**

Configuration–approach pairs for which W(z) = 1, $I_a(z) = 1$ and $U_a(z) = 1$. More than one approach may survive; which to take is the researcher's decision, taken after a named human review.

**Figure 2.** The Conditions for Entering the Theoretical Opportunity Space. Notes. Conditions are applied in the order shown, and the third column names the research action a refusal produces, so a refusal is a recorded routing outcome rather than a null result. The three shaded rows are the evidentiary and diagnostic preconditions shared by all four detectors; the three unshaded rows are the conditions in the definition of $\Omega$. $I_{\mathrm{a}}(z)$ refuses no configuration — it selects which approaches remain eligible. The panel at the foot is the outcome of the chain, not a further condition.

Only configurations with $W(z) = 1$ proceed to theorizing-approach routing. Let $I_{\mathrm{a}}(z)$ indicate whether theorizing approach $a$ is eligible for configuration $z$ under the literature-grounded diagnosis–action crosswalk. Eligibility follows from the diagnosed theoretical problem rather than from novelty or the availability of an unexplored relationship. For example, mechanism-based theorizing is eligible when a sufficiently supported phenomenon or relationship lacks an adequate generative explanation, whereas boundary-condition theorizing is eligible when an otherwise supported explanation changes or fails under theoretically meaningful conditions. The corresponding criteria for deductive elaboration and context-based theorizing are reported in the Online Appendix; the complete routing crosswalk is shown in Figure 7. Multiple approaches may remain eligible when more than one theorizing logic fits the diagnosed problem.

The crosswalk is encoded ex ante rather than selected by the model during a run. Its mappings are derived from the theoretical problem–response logics represented in the focal theorizing approaches and translated into explicit diagnosis–action–approach rules. The Online Appendix documents the derivation of each diagnosis, its associated research action, the theorizing approaches it licenses, and the scholarly sources supporting that mapping. Thus, routing reflects an inspectable design artifact rather than an unconstrained model judgment.

Finally, the platform evaluates prior-art unresolvedness, $U_{\mathrm{a}}(z) \in \{0, 1, \perp\}$ where 1 indicates that the corresponding theoretical issue remains unresolved, 0 that prior theoretical treatment has been identified, and $\perp$that unresolvedness cannot yet be determined. A configuration enters the theory-oriented opportunity set only when theory development is warranted, approach $a$is eligible, and the theoretical issue remains unresolved: $O^{\mathrm{a}}(z) = \mathbf{1}[W(z) = 1]\ I_{\mathrm{a}}(z)\ \mathbf{1}[U_{\mathrm{a}}(z) = 1]$, and $\Omega = \{(z, a) : O^{\mathrm{a}}(z) = 1\}$. Because $U_{\mathrm{a}}(z)$ is evaluated last and may return $\perp$, we name the set the first two conditions define on their own. Let $\Omega_C = \{(z, a) : W(z) = 1 \ and \ I_{\mathrm{a}}(z) = 1\}$ denote the theory-development candidate set: configurations for which theory development is the warranted research action, paired with each approach the crosswalk licenses. By construction $\Omega \subseteq \Omega_C$ , and $\Omega = \{(z, a) \in \Omega_C : U_{\mathrm{a}}(z) = 1\}$. When $U_{\mathrm{a}}(z) = \perp, \mathrm{O}^{\mathrm{a}}(\mathrm{z}) = \perp$ and the pair's membership in $\Omega$ is undetermined rather than refused; the platform records it in $\Omega$_C together with the outstanding condition. The distinction matters because an unassessed condition is not a failed one, and a system that reports theory-development candidates as though they were opportunities has collapsed exactly the escalation this architecture exists to keep separate.

Figure 2 states these conditions as an ordered chain. A configuration is evaluated against each in turn, and every condition that refuses names the research action taken in place of theory development, so a refusal is a recorded routing outcome rather than a null result. Membership in $\Omega$ identifies a theory-oriented opportunity for further evaluation rather than establishing a theoretical contribution. The platform therefore produces evidence-linked research-action and contribution-potential recommendations; final judgments about theoretical contribution remain with the researcher.

### 3.2. Platform Architecture and Three-Stage Design

The platform operationalizes the theoretical-opportunity process through the three-stage architecture in Figure 1, guided by four design principles. First, it follows minimum necessary agency: LLM-based agents are used for tasks requiring open-ended semantic interpretation, discovery, or reconstruction, whereas deterministic components govern explicit state transitions, thresholds, routing rules, and validation procedures (Pacella et al. 2026). Second, the platform enforces epistemic separation across discovery, evidence reconstruction, and research-action judgment. What may be relevant, what the scholarly record supports, and what research action is justified remain distinct states, reducing the risk that generated candidates or downstream interpretations anchor later evidentiary judgments (Nguyen 2024). Third, the architecture uses specialized reasoning with controlled integration. Search, extraction, appraisal, identity resolution, theory reconstruction, and diagnosis are separated when they require distinct analytical logics, while explicit resolvers and cross-stage handoffs govern when outputs can be combined (Li et al. 2023; Chen et al. 2024; Prasad et al. 2024). Fourth, the platform requires verification before epistemic escalation and bounds autonomy through stopping, refusal, and human-governance conditions (Zhu et al. 2026; Nguyen et al. 2026). Candidates advance only when evidentiary, quality, and eligibility conditions are satisfied; incompleteness and non-theory routes remain valid outcomes.

These principles are implemented differently across the three stages. Stage 1 is agent-intensive because discovery requires broad exploration; Stage 2 combines model-based semantic reconstruction with stronger verification and information-isolation controls; and Stage 3 is predominantly deterministic because it converts a frozen evidence-grounded theory state into a research-action recommendation. The architecture therefore separates possibility, evidence, and justified action not only as workflow stages, but as distinct epistemic functions.

### *3.2.1. Stage 1: Discovery Stage*

Stage 1 addresses an open-ended exploration problem: the relevant conceptual space is not yet known, and premature constraints may exclude important theoretical directions. Consistent with minimum necessary agency, Stage 1 therefore makes the greatest use of model-based agents while treating all outputs as provisional candidates rather than evidence-grounded claims. Specialized discovery functions independently explore focal concepts, nomological relations, contexts and processes, and potentially missing connections across weakly connected theoretical regions. Their outputs are integrated through controlled identity resolution, coverage assessment, and routing rules that distinguish literature-linked candidates, search hypotheses not yet located, and novel theory candidates proposed to fill gaps. A separate saturation verdict records whether discovery is complete, budget-limited, or unreliable. Stage 1 produces the Candidate Knowledge Graph and a validation package for Stage 2; detailed agent roles and handoffs are reported in the Online Appendix. The present study evaluates Stage 1 as a high-recall exploration component within an epistemically separated architecture; it does not test whether Stage 1 provides incremental discovery value relative to evidence reconstruction performed without a prior discovery stage.

### *3.2.2. Stage 2: Evidence Reconstruction Stage*

Stage 2 reconstructs what the scholarly record supports while preventing the Stage 1 graph from anchoring source interpretation. Stage 1 may influence source retrieval, but eligible papers are interpreted independently, and reconciliation occurs only after the literature-derived representation has been internally resolved. Retrieval, screening, source-depth classification, source-grounded extraction, study-design appraisal, and quality control are therefore separated from Stage 1 candidate generation. Construct identity, measurement relations, theory architecture, assumptions, mechanisms, and theory evolution are resolved before the Evidence-Adjudicated Knowledge Graph is compared with $G^C$. Reconciliation then classifies Stage 1 candidates as supported, contradicted, unsupported, or unresolved, while preserving literature-supported knowledge that Stage 1 did not anticipate. Remaining incompleteness is recorded rather than filled through inference. Stage 2 passes the resolved evidence graph, reconciliation state, theory state, provenance, uncertainty, and completion status to Stage 3. Detailed modules and controls are reported in the Online Appendix.

#### *3.2.3. Stage 3: Opportunity Diagnosis Stage*

Stage 3 determines whether an unresolved pattern warrants theory development or another research action. The current implementation is intentionally non-generative: it makes no model calls and uses deterministic rules for evidence thresholds, diagnosis, routing, and governance. Phase 1 constructs a frozen Evidence-Adjudicated Theory State through EATSView and SyncController. Stage 3 can reason from this state but cannot repair missing evidence or write downstream interpretations back into it. Phases 2 and 3 then use SignalTriageController and VerificationController to identify and verify focal configurations. Conceptual-clarity and identity checks prevent unverified patterns from entering theory-oriented action, with ambiguous identities escalated to human adjudication.

The evaluated kernel diagnoses evidence of adequacy and four kinds of theoretical problem: a missing mechanism, a boundary failure, a contextual failure, and a misspecified relationship. Each detector shares one precondition — the relationship must be evidentially established — and then applies its own. A single configuration may therefore present more than one typed problem, and the crosswalk maps each problem, not the configuration, to the theorizing approaches it licenses.

Phase 4 converts diagnosis into a research-action recommendation through TheoryActionGate and ResearchActionPlanController. Sixteen action types are evaluated against five tests, and theory generation is blocked when no action record passes. Additional evidence, replication, measurement work, conceptual clarification, causal identification, method or data development, empirical-scope work, and no theory development are therefore legitimate outcomes. Only when theory development is warranted does the system determine which theorizing approach is eligible. The broader repertoire library supports additional modes, but the present study evaluates only the focal approaches specified in the theoretical framework. Phase 5 tests surviving candidates against prior art, counterevidence, and competing explanations under explicit budgets. The current implementation does not generate a missing theorizing move: if no external candidate is supplied, the run stops rather than inventing one. If a candidate is supplied, Stage 3 can evaluate it using published evidence.

Phase 6 establishes the final boundary between system recommendation and scholarly authority. HumanAIConvergenceGate requires named human review for Tier A or B outcomes. Final outputs include configuration–approach pairs that enter the theoretical opportunity space, non-theory research routes, and refused candidates with recorded attrition reasons. The system therefore structures and constrains recommendations, while the researcher retains authority over whether a theoretical opportunity is accepted and how any resulting contribution is claimed.

# 4. Empirical Results

## 4.1. Data and Corpus Construction

To evaluate the platform, we conducted an end-to-end empirical run using *Human Oversight of Agentic AI Systems* in Organizations as the focal research topic. Because the three stages serve different epistemic purposes, the scholarly material used by the platform was assembled progressively rather than defined as a single fixed corpus at the outset. Table 3 summarizes the principal data states used across the Discovery Stage and the Evidence Reconstruction Stage.

**Table 3.** Principal Data States Across the Discovery and Evidence Reconstruction Stages

| **Stage** | **Corpus / data state** | **n** |
|---|---|---|
| Stage 1: Discovery Stage | Scholarly works retrieved for candidate discovery | 1,066 |
| | Initial retrieved literature pool | 7,028 |
| Stage 2: Evidence Reconstruction Stage | Works retained for evidence reconstruction and adjudication | 1,415 |
| | Works available at ≥ partial full-text depth | 729 |

The Evidence Reconstruction Stage began from a broader pool of 7,028 unique scholarly works assembled through independent topic retrieval, targeted retrieval, and citation-based expansion. Following screening, 1,415 works were retained for evidence reconstruction and adjudication. Because the strength of downstream claims depends on source accessibility, the platform separately tracked evidence depth, with 729 works available at least at partial full-text level. These successive corpus states reflect increasing evidentiary depth within Stage 2 rather than separate independent datasets.

## 4.2. Results

### *4.2.1. Discovery Produces a Broad but Predominantly Unverified Possibility Space*

The Discovery Stage produced a Candidate Knowledge Graph ($G^c$) containing 7,557 candidate nodes and 6,585 proposed relationships. In this graph, a candidate node represents a construct that the discovery agents identify as potentially relevant to the focal theoretical space, while a proposed relationship represents a possible theoretical connection between two such constructs. These objects are provisional rather than evidence-grounded: they may be recalled from prior scholarly knowledge, inferred as search hypotheses, or generated as possible theoretical extensions. Stage 1 therefore represents a broad space of theoretical possibilities rather than a summary of what the literature has established.

Because the same relationship may be proposed through multiple agents or search paths, the graph distinguishes individual proposals from consolidated relationships. Its largest connected component contains 3,382 candidate construct senses joined by 5,767 distinct proposed relationships, representing 6,523 separate proposals; 2,325 of those senses appeared in exactly one relationship. Figure 3 shows that the resulting search space is organized around several broad conceptual regions, including trust and reliance, human oversight and workload, meaningful control and accountability, under-reliance and dispositional trust, and algorithmic management. These regions are connected through a smaller number of high-degree or bridging constructs, while more specialized candidates extend toward the periphery. The pattern shows that discovery produced a broad and structured possibility space, but neither a construct's location nor its connectivity should be interpreted as evidence of theoretical importance or empirical support.

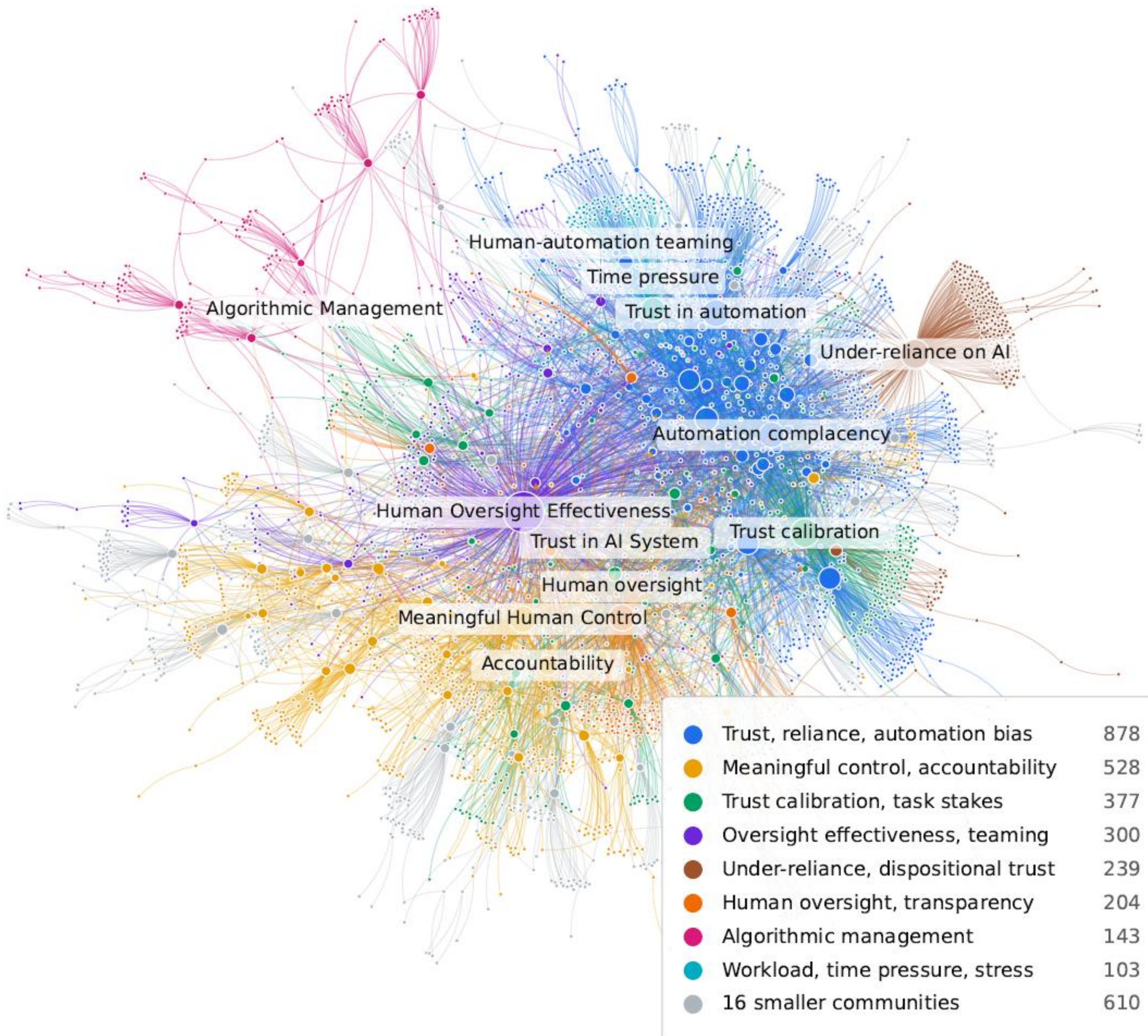

**Figure 3.** Stage 1 Candidate Knowledge Graph. *Notes*. The largest connected component of $G^{c}$. Color is graph community; the legend names the eight largest by their highest-degree members and gives the size of each, with the remaining 16 pooled in grey. Node area is proportional to degree, subject to a floor below roughly degree 5. Link width is the number of independent proposals naming the pair, and an arrowhead appears where those proposals agree on a direction. Labels mark each community's highest-degree construct, and the top two where a community exceeds 300 members.

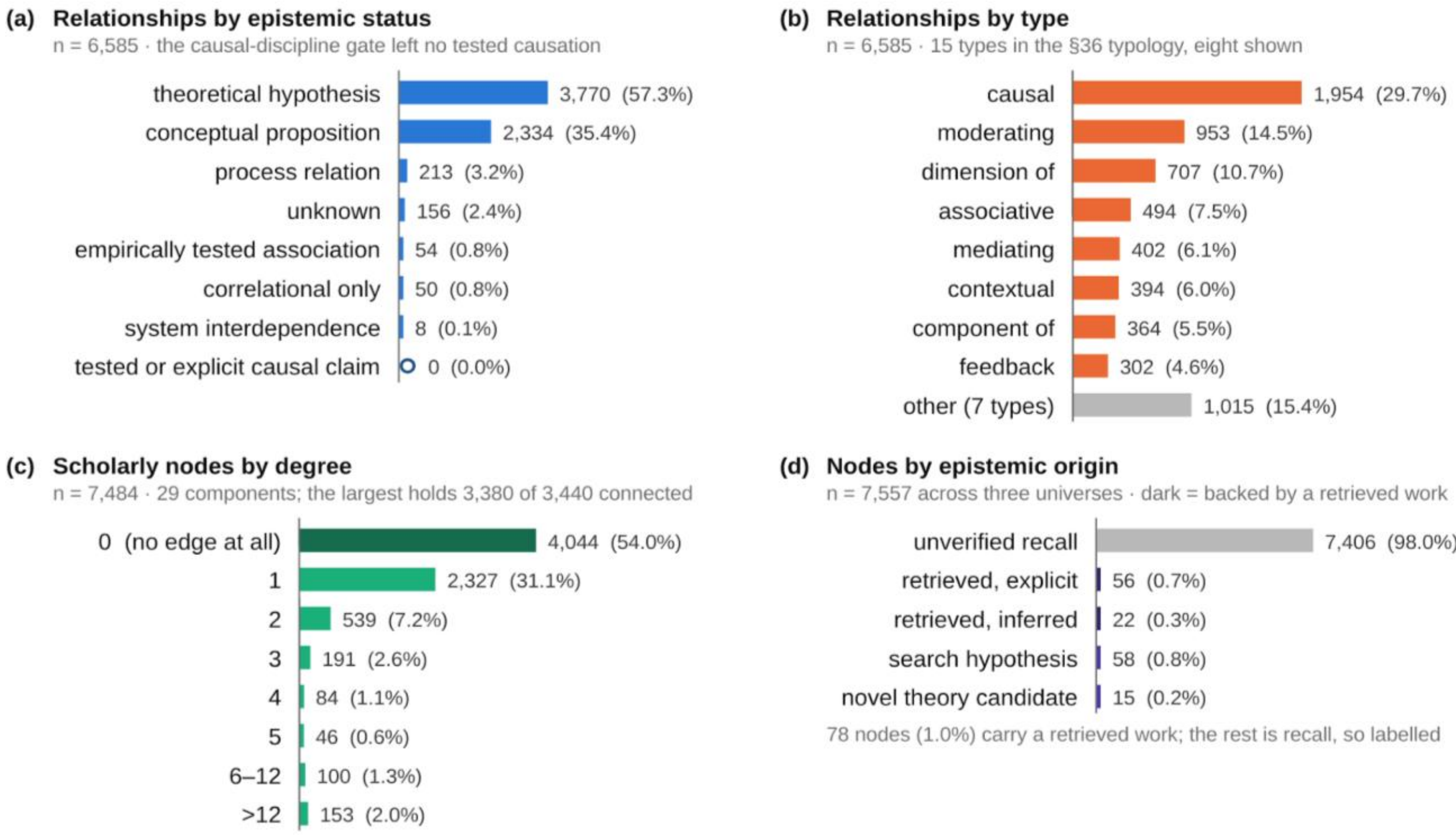

**Figure 4.** Epistemic and Structural Composition of $G^c$. Notes. Panels (a) and (b) describe the 6,585 proposed relationships, (c) the 7,484 scholarly candidate nodes, and (d) all 7,557 candidate nodes; the three bases are distinct populations, not nested subsets. In (a), the open marker records a category with zero members. In (b), eight of the fifteen relationship types are shown individually and the remaining seven are pooled. In (c), degree is computed within $G^c$, so a node at degree 0 was proposed with no relationship at all. Color distinguishes panels and encodes nothing.

Figure 4 clarifies the epistemic and structural composition of $G^c$. The proposed relationships (Figure 4a) were dominated by theoretical hypotheses (57.3%) and conceptual propositions (35.4%), while only 0.8% were labeled empirically tested associations and none were classified as tested or explicit causal claims. At the same time, the candidate graph contained multiple relationship types (Figure 4b) , indicating that discovery was not restricted to a single form of theoretical connection. Structurally, in Figure 4c, 54.0% of scholarly candidate nodes had no edge, indicating that the breadth of the graph partly reflected the retention of isolated candidates rather than only densely connected constructs. Most importantly, in Figure 4d, 98.0% of Stage 1 nodes were explicitly labeled as unverified recall, whereas only 1.0% were backed by a retrieved scholarly work. This distribution reflects the **breadth-oriented design** of the Discovery Stage rather than a claim that these candidates represent established knowledge. Stage **1** expands the candidate space while preserving the distinction between generated possibilities and independently verified scholarly evidence**.**

#### *4.2.2. Evidence Reconstruction Substantially Changes the Knowledge State*

Stage 2 independently reconstructed the scholarly record from eligible papers at the available full- or partial-text depth, without using the Stage 1 candidate graph $\boldsymbol{G}^{\mathrm{c}}$ as an evidence base or extraction template**.** Whereas Stage 1 generates candidate constructs and relationships, Stage 2 determines what construct senses, claims, and relationships can be reconstructed from retrieved literature**.** The resulting Evidence-Adjudicated Knowledge Graph ($G^{\mathrm{E}}$) contained 18,203 evidence-grounded construct nodes with at least one supporting claim and 14,214 adjudicated relationships derived from 16,414 source-grounded claims. A relationship is included in $G^{\mathrm{E}}$ when at least one source-grounded claim supports its presence in the literature. However, inclusion in the graph does not mean that the relationship is already well established.

The resulting graph therefore represents a qualitatively different knowledge state from the Stage 1 candidate graph. *Trust*, *transparency*, *explainability*, *automation bias*, *human oversight*, *AI adoption*, and related constructs remain visible, but they are organized around construct senses resolved from the literature rather than around generated candidate labels. Stage 2 also retains information needed to interpret these relationships, including whether similarly labeled constructs refer to the same concept, how relationships are supported across sources, how constructs are measured, and what theoretical conditions or mechanisms are associated with them.

The final step of Stage 2 compares the independently reconstructed $G^{\mathrm{E}}$ with the candidate space generated in Stage 1. The overlap between the two representations was limited. At the construct level, 2,689 of the 7,484 Stage 1 scholarly candidates (35.9%) could be matched to a corresponding construct sense in $G^{\mathrm{E}}$. The overlap was substantially smaller at the relationship level, with almost none of the specific relationships proposed in Stage 1 recovered under strict construct-sense matching. These results illustrate that the possibilities generated during discovery can differ substantially from the relationships reconstructed from the scholarly record once source and construct-identity constraints are imposed.

The limited overlap between $G^{\mathrm{c}}$ and $G^{\mathrm{E}}$ should not itself be interpreted as validation of the architecture, as it may also reflect speculative generation, incomplete retrieval, or unresolved construct identity. The reconstructed $G^{\mathrm{E}}$ provides the evidence base for Stage 3, which evaluates whether any of these literature-grounded relationships are sufficiently established to warrant theoretical diagnosis.

#### *4.2.3. Most Literature-Grounded Relationships Do Not Warrant Theory Development*

Stage 3 evaluates whether the literature-grounded relationships reconstructed in Stage 2 are sufficiently established to support theoretical diagnosis. Relationships that lack adequate multi-study evidence or directional consistency are routed to additional evidence or replication rather than theory development. Only relationships that pass these evidentiary gates proceed to diagnosis and theorizing-approach routing; final opportunity-space membership additionally requires prior-art unresolvedness. Stage 3 therefore treats theoretical opportunity as an adjudicated outcome rather than a graph gap.

Figure 5 shows the resulting attrition. Of the 14,214 reconstructed relationships, only 28 were supported by at least three independent works, and only two satisfied the directional-establishment criterion. The direction test assigns one vote per independent work when that work reports an observed signed relationship. Transparency → trust received 10 directional votes from 14 works, all positive, while explainability → trust received six directional votes from nine works, also all positive. These two relationships therefore proceeded to theoretical diagnosis; the remaining relationships did not proceed because the current evidence state did not satisfy the evidentiary gates.

Although the focal domain is human oversight of agentic AI systems in organizations, both retained relationships originate in the broader human-AI literature. They are retained because transparency and explainability are theoretically relevant to how users evaluate and rely on AI-mediated action. Their inclusion therefore reflects theoretically adjacent knowledge rather than evidence that the underlying studies directly examined contemporary agentic AI systems.

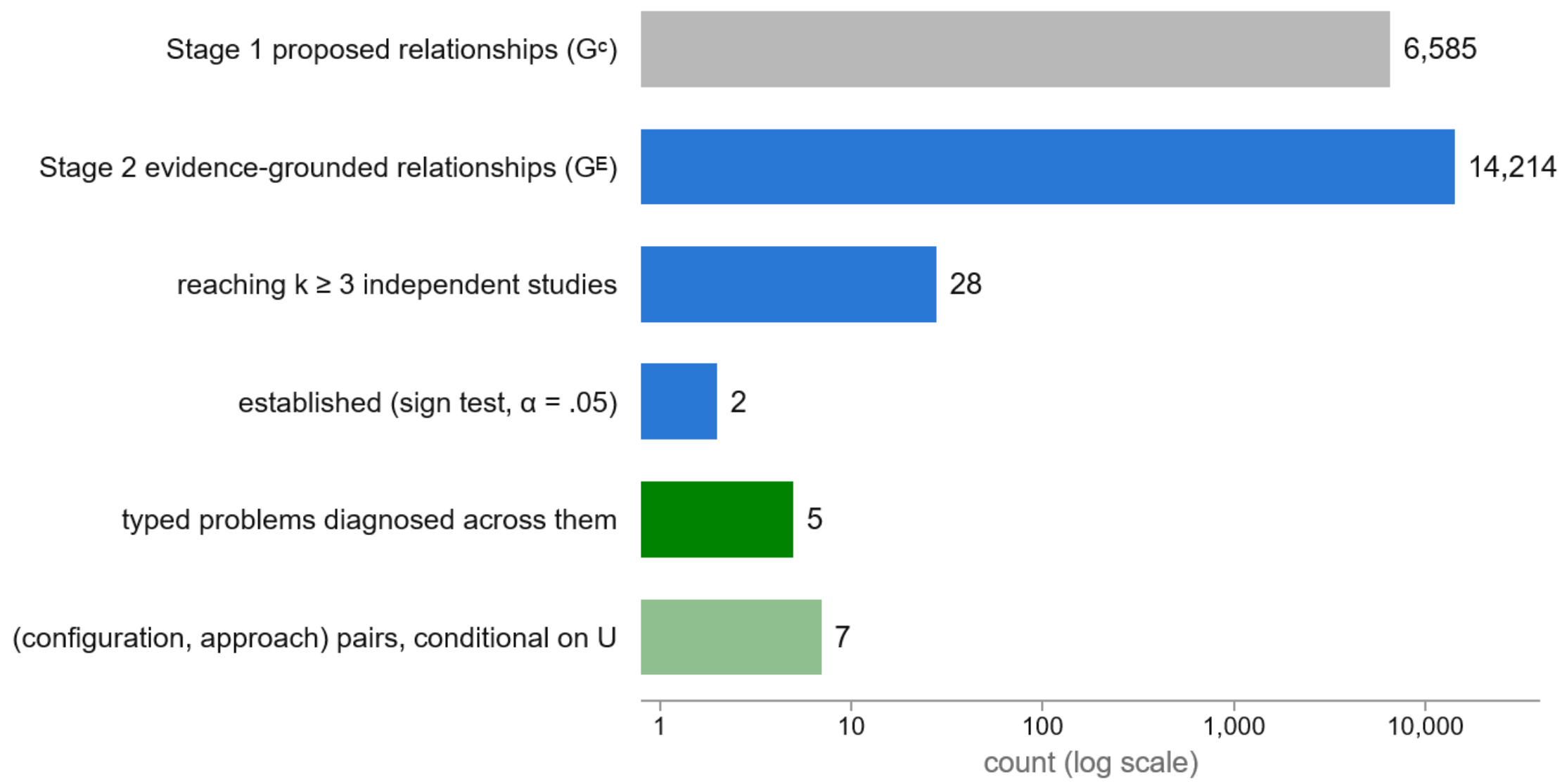

**Figure 5.** Progressive Attrition Toward the Theoretical Opportunity Space. Notes. Counts on a logarithmic scale. The first bar is shown for scale only: $G^{\mathrm{c}}$ is a different graph, not a superset of $G^{\mathrm{E}}$. Bars two through four are nested. The last two bars count typed diagnoses and (configuration, approach) pairs rather than relationships, so they are not a reversal of the attrition; Section 4.2.3 explains the step. $U_{\mathrm{a}}(z)$ was not assessed, so these are members of the candidate set $\Omega_C$; membership in $\Omega$ is undetermined.

Figure 6 shows the diagnoses and resulting theorizing routes for the two retained relationships. Transparency → trust presents three typed problems: boundary failure, missing mechanism, and a misspecified relationship; explainability → trust presents boundary failure and missing mechanism. Applying the fixed diagnosis–action–approach crosswalk produces seven distinct configuration–approach pairs. Both relationships are eligible for mechanism-based, boundary-condition, and context-based theorizing, while transparency → trust is additionally eligible for deductive elaboration. Because multiple diagnoses can license the same approach for the same configuration, the five diagnoses generate nine routing records but only seven distinct pairs. These seven pairs therefore constitute the theory-development candidate set $\Omega_C$ for this run. Each satisfies the theory-development warrant and approach-eligibility conditions, $W(z) = 1$ and $I_{\mathrm{a}}(z) = 1$. Because prior-art unresolvedness, $U_{\mathrm{a}}(z)$, was not assessed, their membership in the final theoretical opportunity space, $\Omega$, is undetermined.

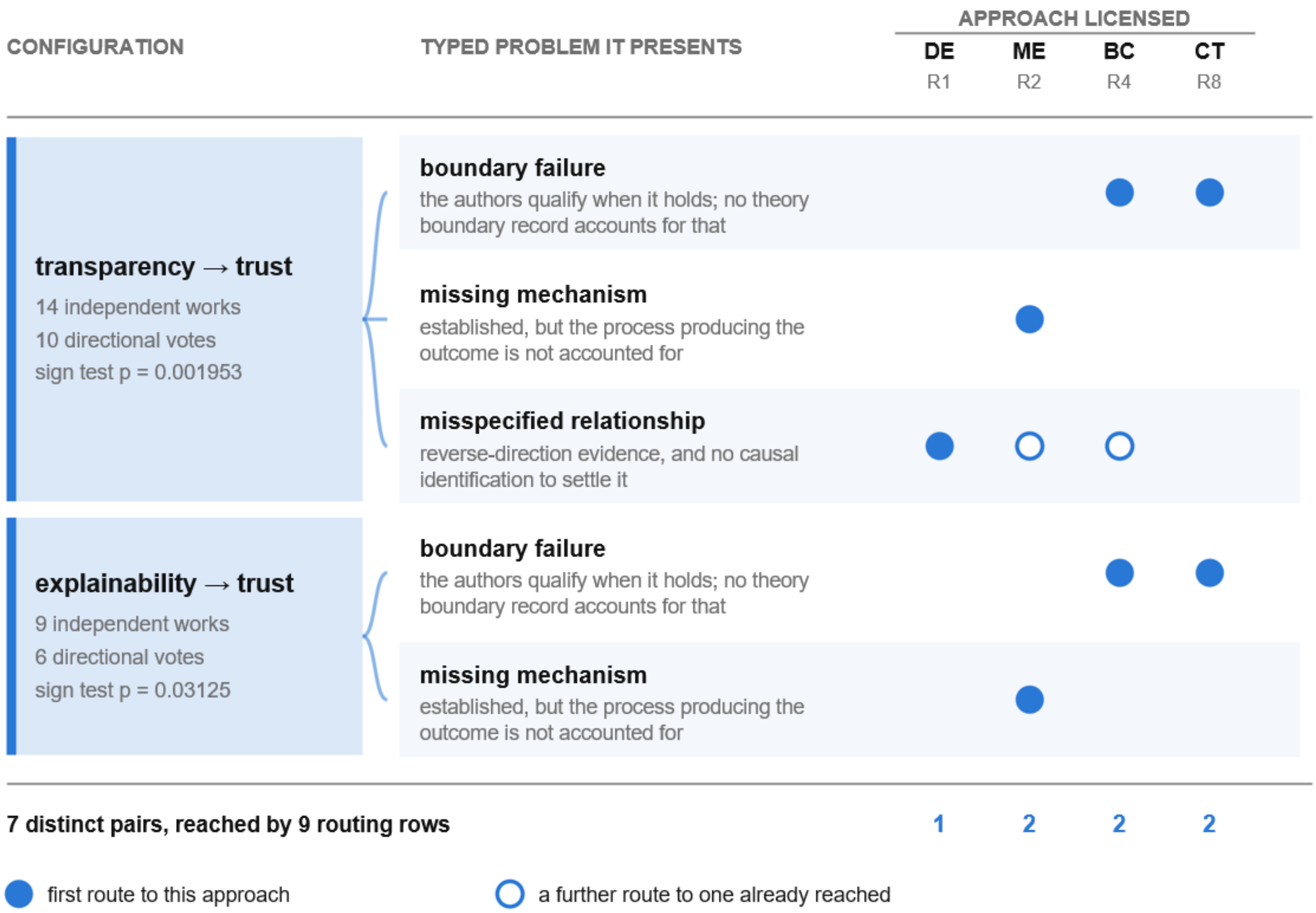

**Figure 6.** How the Two Admitted Configurations Reach Their Theorizing Approaches. Notes. Each row is one (configuration, typed problem) pair, with the bracket joining a configuration to the problems its evidence state presents. R1, R2, R4 and R8 are the platform's repertoire identifiers for the four approaches of Figure 6. A filled disc is the first route by which a configuration reaches an approach; an open ring is a further route to one already reached — which is why nine routing rows yield seven distinct pairs. The foot row counts distinct pairs per approach. Directional votes are the independent works that state a direction; all agreed in sign.

These results should be interpreted within the limits of the current run. Most detector evaluations stopped at the shared relationship-establishment gate, so the run demonstrates the operation of the routing logic on the admitted configurations rather than characterizing detector behavior at scale. The run was also incomplete rather than saturated, and some mechanism and boundary records could not be linked to reconstructed relationships. Accordingly, an apparent absence may reflect either a genuine theoretical deficiency or incomplete extraction and linkage. Section 4.3 examines the robustness and traceability of the resulting Stage 3 decisions.

### 4.3. Routing Evaluation and Robustness Checks

We evaluate the current Stage 3 decisions through three complementary checks: contrastive discrimination among nearby relationships, traceability of diagnoses and theorizing routes to the reconstructed evidence state, and robustness to alternative evidence thresholds and corpus restrictions. These checks assess the internal coherence, traceability, and robustness of the evidence-to-diagnosis-to-action process; they do not constitute independent expert validation of the underlying literature. The routing criteria determine whether the reconstructed evidence state is sufficiently developed to permit theoretical diagnosis; they do not estimate effect magnitude, establish causality, or demonstrate scientific consensus.

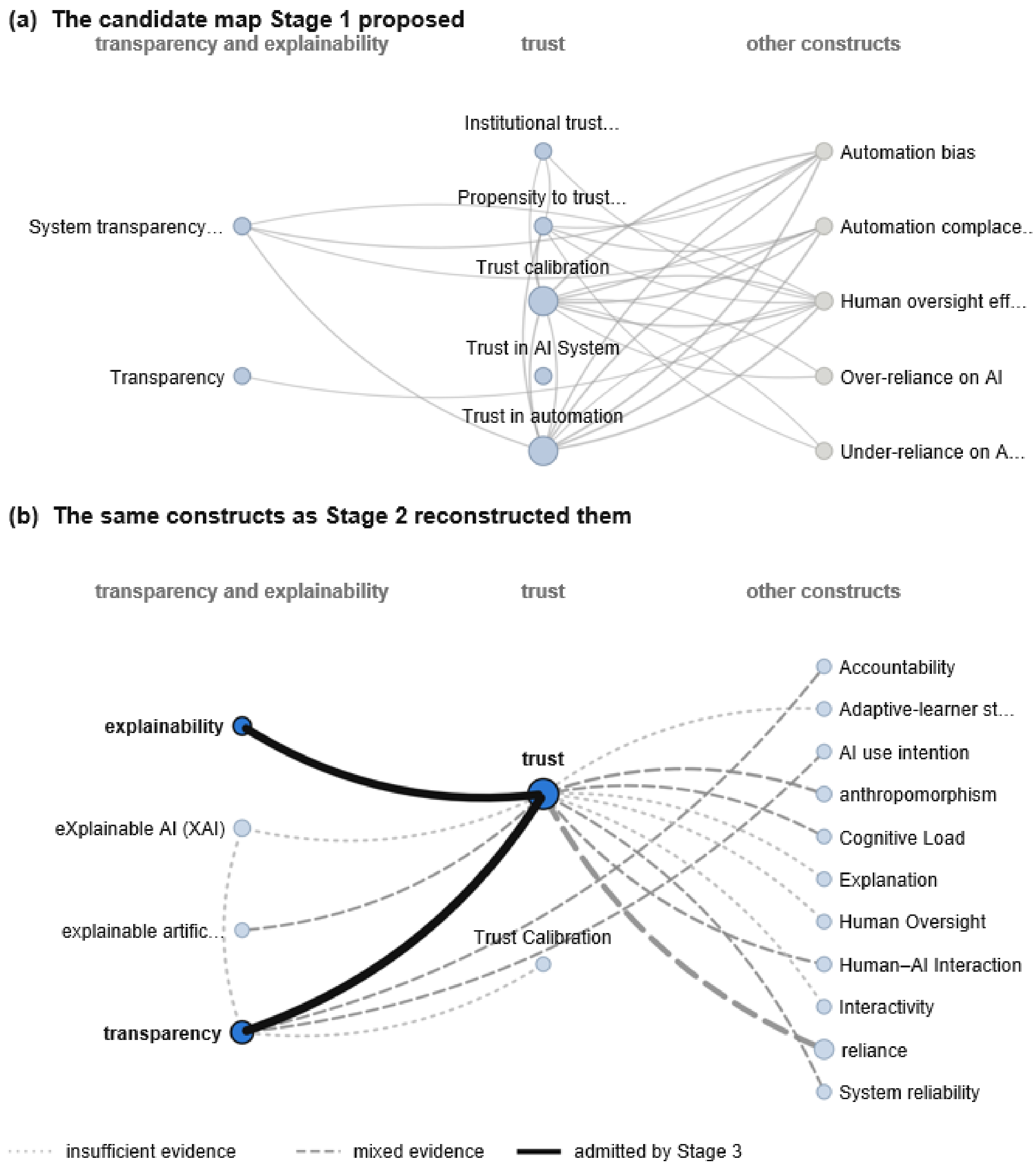

**Figure 7.** Contrastive Validation Within the Transparency–Explainability–Trust Neighborhood. Notes. The same region of literature as Stage 1 proposed **(a)** and as Stage 2 independently reconstructed it **(b)**. Constructs sit in three columns — transparency and explainability, trust, other constructs — and vertical position within a column carries no meaning. Node area grows with degree in **(a)** and with the number of adjudicated claims in **(b)**; link width grows with claims. Line style and darkness encode Stage 2's adjudication status, which is ordinal and so is not encoded by hue; a heavy solid black link is a relationship Stage 3 admitted. Panel **(b)** is filtered for legibility, showing only the 17 relationships that carry four or more claims out of 472 in the neighborhood. Labels are truncated at 20 characters.

As shown in Figure 7, the local neighborhood provides a contrastive check on whether Stage 3 distinguishes among nearby relationships rather than treating evidence volume alone as sufficient for theory development. Trust → reliance contained 20 claims from 10 sources but failed the directional criterion and was routed to MORE_EVIDENCE_FIRST, whereas transparency → accountability was routed to REPLICATION_FIRST because its evidence was directionally conflicting. Only transparency → trust and explainability → trust passed both the evidence and directional gates and proceeded to theoretical diagnosis.

For those two admitted relationships, the theorizing routes are traceable to specific deficiencies in the reconstructed theory state. Neither configuration contained a sufficiently specified generative mechanism, supporting a missing-mechanism diagnosis, and both contained author-stated scope qualifiers not accounted for by a linked theory-boundary record, supporting a boundary-failure diagnosis. Transparency → trust additionally carried reverse-direction evidence without causal identification, producing a misspecified-relationship diagnosis. The resulting theorizing routes therefore follow from the reconstructed evidence state rather than from relationship prominence or evidence volume alone.

Table 4 reports sensitivity to alternative evidence thresholds. Under the stricter profile ($k \geq 5$, $\alpha = .01$), transparency → trust remains admitted but explainability → trust does not. Under the more permissive profile ($k \geq 3$, $\alpha = .10$), no additional relationship enters the admitted set. The substantive conclusion is therefore stable for transparency → trust but threshold-sensitive for explainability → trust.

**Table 4.** Threshold sensitivity of Stage 3 theory-development routing.

| **Profile** | **k** | **α** | **Reached k** | **n retained** | **Retained relationships** |
|---|---|---|---|---|---|
| Strict | ≥ 5 | .01 | 4 | 1 | Transparency → trust |
| Default (TP-1.0) | ≥ 3 | .05 | 28 | 2 | Transparency → trust; Explainability → trust |
| Permissive | ≥ 3 | .10 | 28 | 2 | Transparency → trust; Explainability → trust |

A second robustness check narrows the source pool to the 1,415 works retained in the adjudication corpus. Because extraction was conducted across both discovery and adjudication corpora, 40.3% of the 16,414 claims came from discovery-only sources. Under the restricted corpus, the number of relationships meeting the multi-work evidence threshold falls from 28 to 15 and the established set falls from two to one. Transparency → trust remains established with a 7–0 directional split (p = .0156), whereas explainability → trust falls to three directional works and cannot satisfy the default two-sided significance criterion. This result does not imply that extraction from discovery-only sources was inappropriate; rather, it shows that explainability → trust is more sensitive to inclusion of the broader source pool, whereas transparency → trust is not.

Taken together, these checks validate the coherence and traceability of the current Stage 3 routing logic. They do not establish that the published literature definitively lacks mechanisms for these relationships: the mechanism diagnosis is conditional on the reconstructed evidence state, and direct source-level human adjudication has not yet been completed. Accordingly, the present empirical claim stops at theory-development warrant and approach eligibility. Completing $U_{\mathrm{a}}(z)$ is what would resolve $\Omega \subseteq \Omega_C$; until then the run's terminal output is $\Omega_C$.

## 5. Discussion and Future Work

This study develops a theory-oriented AI research assistant designed to help researchers identify potential theorizing opportunities from large bodies of scholarly literature. The platform operates through three stages with distinct functions. Stage 1 broadly discovers potentially relevant constructs and relationships, Stage 2 independently reconstructs the evidence and theoretical state represented in the literature, and Stage 3 evaluates which unresolved configurations warrant theory development and routes them to appropriate theorizing approaches. Importantly, these opportunities are not generated as open-ended research ideas or inferred simply from gaps or missing links in the literature. Instead, their identification is guided by established approaches to theorizing, including deductive elaboration, mechanism-based theorizing, boundary-condition theorizing, and context-based theorizing. By combining AI's capacity to process large volumes of scholarly evidence with structured knowledge representation and theory-guided diagnosis, the platform can help researchers focus on theoretically meaningful directions that provide a stronger basis for subsequent human inquiry and theory development.

Several limitations remain. The empirical evaluation covers a single focal topic, and although all four focal theorizing routes were reached, each rests on one or two configurations, so the run demonstrates that the routing operates rather than establishing how it behaves at scale. In addition, the present validation focuses on the coherence, traceability, and robustness of the routing process rather than full human source-level adjudication, and unresolvedness $U_{\mathrm{a}}(z)$ was not completed for the two retained configurations. Future work should therefore evaluate the platform across additional domains and theorizing routes, improve full-text and theory-state coverage, and compare system judgments more directly with expert human assessments.